\documentclass[preprint,12pt]{elsarticle}

\usepackage{amssymb}
\usepackage{lineno}

\newcommand{\ddt}[1]{{\frac{\partial {#1}}{\partial t}}}
\newcommand{\ddx}[1]{{\frac{\partial {#1}}{\partial x}}}
\newcommand{\ddvx}[1]{{\frac{\partial {#1}}{\partial v_{x}}}}

\newcommand{\vv}{{{\bf v}}}

\newcommand{\inv}[1]{{\frac{1}{#1}}}

\newcommand{\etal}{{{\it et. al.}}}
\newcommand{\eg}{{{\it e.g.}}}

\journal{Planetary and Space Science}

\begin{document}

\begin{frontmatter}



\title{Nonlinear evolution of parallel propagating Alfv\'en waves: Vlasov - MHD simulation}


\author[label1]{Nariyuki, Y.}
\author[label2]{Umeda, T.}
\author[label3]{Kumashiro, T.}
\author[label3]{Hada, T.}

\address[label1]{Department of Electrical Engineering and Information Science, Kochi National College of Technology, Kochi 783-8508, Japan}
\address[label2]{Solar-Terrestrial Environment Laboratory, Nagoya University, Nagoya, Aichi 464-8601, Japan}

\address[label3]{E.S.S.T., Kyushu University, 6-1, Kasuga-koen, Kasuga City, Fukukoka 816-8580, Japan}
\begin{abstract}
Nonlinear evolution of circularly polarized Alfv\'en waves are discussed by using the recently developed Vlasov-MHD code, which is a generalized Landau-fluid model. The numerical results indicate that as far as the nonlinearity in the system is not so large, the Vlasov-MHD model can validly solve time evolution of the Alfv\'enic turbulence both in the linear and nonlinear stages. The present Vlasov-MHD model is proper to discuss the solar coronal heating and solar wind acceleration by Alfve\'n waves propagating from the photosphere.

\end{abstract}


\end{frontmatter}


\linenumbers 
\section{Introduction}
Large-amplitude Alfv\'enic fluctuations are ubiquitous in the heliosphere, especially in the fast solar wind (Bruno and Carbone, 2005). Nonlinear evolution and dissipation of these fluctuations are thought to play important roles in heating of the solar wind plasmas (Suzuki and Inutsuka, 2006; Wu and Yoon,2007; Araneda \etal, 2008; Valentini \etal, 2008) and generation of the localized structures (Tsurutani \etal, 2005; Vasquez \etal, 2007; Lin \etal, 2009). Since these fluctuations are typically robust for linear ion-cyclotron damping due to their small wave frequencies and for linear Landau damping due to their small propagation angle relative to the background magnetic field, wave-wave interactions (parametric instabilities) are significant processes of their nonlinear evolution. 

The quasi-parallel propagating Alfv\'enic fluctuations can resonate with both the parallel and obliquely propagating magnetohydrodynamic (MHD) waves (Mjolhus and Hada, 1990; Champouex \etal, 1999; Nariyuki \etal, 2008). In uniform plasmas with a typical parameter of the solar wind, parallel propagating Alfv\'en waves dominantly resonate with the parallel propagating Alfv\'en waves and ion acoustic waves. It is important that Alfv\'enic fluctuations can resonate with the \lq\lq kinetic'' wave modes such as ion acoustic waves. Namely, even if the fluid approximation is valid for Alfv\'enic fluctuations themselves, the ion kinetics should be considered for parametric instabilities (Araneda, 1998; Bugnon \etal, 2004; Nariyuki and Hada, 2006; 2007, Araneda \etal, 2007).

Furthermore, in the plasmas with inhomoginity and/or unstable velocity distribution functions such as beam components, the resonance and dissipation processes can be different from those in the homogenous plasmas (Tsikrauri \etal, 2005; Suzuki and Inutsuka, 2005; 2006; Wang \etal, 2006; Suzuki, 2008; Nariyuki \etal, 2009). The kinetic Alfven waves which propagate at oblique angles relative to the background magnetic field can be excited by the proton beams in the solar wind (Daughton and Gary, 1998; Yin \etal, 2007). We note that such beam-excited waves can preferentially interact with the finite-amplitude parallel propagating Alfv\'en waves. Actually, the observational studies imply the importance of the kinetic Alfv\'en waves on the dissipation of the Alfv\'enic turbulence (Leamon \etal, 1998; Hamilton \etal, 2008). 

As mentioned above, the nonlinear evolution and dissipation processes of Alfv\'enic fluctuations in the solar wind are \textit{cross scale coupling processes}, in which the MHD-, ion-, and electron-scale phenomena coexist and mutually interact. Moreover, it is worth noting that the heating and acceleration process of plasmas by Alfv\'enic fluctuations are desired to be slot into the heliospheric global simulation models (\eg, Nakamizo \etal, 2009). Thus, the development of the alternative \lq\lq kinetic'' MHD models, which are the \lq\lq middel scale'' model including some non-MHD effects, is necessary for \textit{systematic} understanding on such cross scale coupling processes. 

As one of such models, we have recently developed a new Vlasov simulation code named Vlasov-MHD code (1-D in the configuration space, 1-D in \lq\lq kinetic '' velocity space, and 2-D in \lq\lq MHD '' velocity space), in order to study basic properties of nonlinear evolution of Alfv\'en waves (Kumashiro \etal, 2009). The concept of the Vlasov- MHD model is the generalized model of the so-called Landau fluid model (Passot and Sulem, 2003; Bugnon \etal, 2004), which includes the nonlinear wave-particle interactions. The linear analysis of the Vlasov-MHD model has been carried out (Araneda, 1998; Nariyuki and Hada, 2006; 2007, Araneda \etal, 2007) and has concluded that as far as the amplitude of the parent waves is not so large, the growth rates of the linear analysis are consistent with those in the numerical results of the ion hybrid simulation. Kumashiro \etal (2009) performed the numerical simulation using the Vlasov-MHD code with the Hall-effect (Vlasov-Hall-MHD code) and demonstrated that the linear growth of parametric instabilities of Alfv\'en waves are almost consistent with those of the ion hybrid simulation.

In the present paper, we discuss the nonlinear evolution of Alfv\'en waves using the Vlasov-Hall-MHD code. In section 2, we briefly introduce the numerical schemes of Vlasov-Hall-MHD code. We present the simulation results in section 3. Section 4 summarizes the results and briefly discuss the future issues.

\section{Vlasov-Hall-MHD code}
Assuming weak ion cyclotron damping, we include the kinetic effects only along the longitudinal ($x$) direction. Let the ion distribution function $F(x, t ,\vv)$ be separated into the longitudinal and perpendicular directions as follows
\begin{equation}
F(x,t,\vv)=f(t,x,v_{x})g(t,x,v_{y},v_{z}),   \label{eq203}
\end{equation}
the Vlasov - Hall -MHD equations are obtained as follows (Araneda, 1998; Nariyuki and Hada, 2006; 2007)

\begin{equation}
\ddt{f} = - v_x \ddx{f} + \inv{\rho} \ddx{} \left( T_e \rho + \frac{|b_\perp|^2}{2} \right) \ddvx{f} , \label{eq221}
\end{equation}
\begin{equation}
\ddt{u_{\perp}} = - u_{x} \ddx{u_\perp} + \inv{\rho}\ddx{b_\perp} , \label{eq222}
\end{equation}
\begin{equation}
\ddt{b_{\perp}} = \ddx{} \left( u_\perp - u_{x} b_\perp - \frac{i}{\rho}\ddx{b_\perp} \right) . \label{eq223}
\end{equation}
\\
where 
$\rho$ is the plasma density (quasi-neutrality assumed), $u_{x}$ is the longitudinal bulk velocity, $b_{\perp}=b_y + ib_z$ and $u_{\perp}=u_y + iu_z$ are the complex transverse magnetic field and bulk velocity, and $e_x$ is the longitudinal electric field, respectively. All the normalizations have been made using the background constant magnetic field, density, Alfv\'en velocity, and the ion gyro-frequency. 

The total pressure is given as $p=p_{e}+p_{i}$, where $p_{e}$ and $p_{i}$ are electron and ion (proton) pressure, respectively. In the present study, isothermal electrons are assumed ($ p_{e} = T_{e} \rho$) (, namely, the total energy in the system is not conserved). It is also assumed that the ion and electron pressures are isotropic. 

In this study, we solve the Vlasov equation (\ref{eq221}) with the time-advance algorithm called \lq\lq splitting method'' (Cheng and Knorr, 1976), in which the Vlasov equation (\ref{eq221}) is split into the following two advection equations:

\begin{equation}
\ddt{f} = - v_x \ddx{f} \label{eq228}
\end{equation}
\begin{equation}
\ddt{f} = \inv{\rho} \ddx{} \left( T_e \rho + \frac{|b_\perp|^2}{2} \right) \ddvx{f}. \label{eq229}
\end{equation}
The splitting scheme is widely used because of its simplicity of the algorithms and ease of programming. The time advance of distribution function $f(x , v_{x})$ is first carried out by shifting the distribution function in the $x$ direction (\ref{eq228}) with the time step $\Delta t / 2$, shifting the distribution function in the $v_x$ direction (\ref{eq229}) with the time step $\Delta t$ and again shifting the distribution function in the $x$ direction (\ref{eq228}) with the time step $\Delta t / 2$. The spatial profiles of plasma density $\rho$ is computed by integrating the distribution function over $v_x$. In parallel with solving the Vlasov equation, the transverse momentum equation (\ref{eq222}) and the induction equation (\ref{eq223}) are solved by the rational Runge-Kutta scheme for time integration and the spectral method for evaluating spatial derivatives. The number of cells is $1024 \sim 2048$ in the $x$ direction and is $600$ in the $v_{x}$ direction over a velocity range from $v_{max}=2.4$ to $v_{min}=-2.4$. The grid spacing is equal to $\Delta x =0.25$, and the time step is equal to $\Delta t =0.01$. The boundary condition is periodic for the configuration space and the free boundary for velocity space. We adopt PIC scheme (Positive Interpolation for hyperbolic Conservation laws) suggested by Umeda (Umeda, 2008) for time advancement of the Vlasov equation. 

To analyze the parametric instability of Alfv\'en waves, we give monochromatic, circular polarized and parallel propagating parent waves as initial conditions. The initial Alfv\'en wave is written as $b_\perp = b_0 \exp {(-i k_{0} x )}$, $u_\perp = u_0 \exp {(-i k_{0} x )}$, where $b_0$ is the amplitude of parent Alfv\'en waves, $u_0 = - b_0/v_{\phi 0}$ (Walen relation), where the phase velocity $v_{\phi 0} = \omega_0/k_0 (>0)$, $\omega_{0}^{2} = k_{0}^{2} (1+ \omega_{0})$. We adopt the notation that the positive and negative $\omega_{0}$ corresponds to the right hand polarized (RH-) and left hand polarized (LH-) waves. The plasma density is $\rho= \rho_0=1$, $u_{x} = u_{x 0}=0$, and $f=f_0$ is given by the isotropic Maxwellian distribution function. Superposed with the parent wave is a small amplitude white noise with $<|\rho_{noise}|^2>^{1/2}=10^{-5}$.

In the present paper, the ion hybrid simulations are also performed in a way very similar to that described in Nariyuki \etal (2007) to compare the results with those obtained by the Vlasov-MHD simulation.

\section{Simulation results}

\begin{figure}
\noindent\includegraphics[width=22pc]{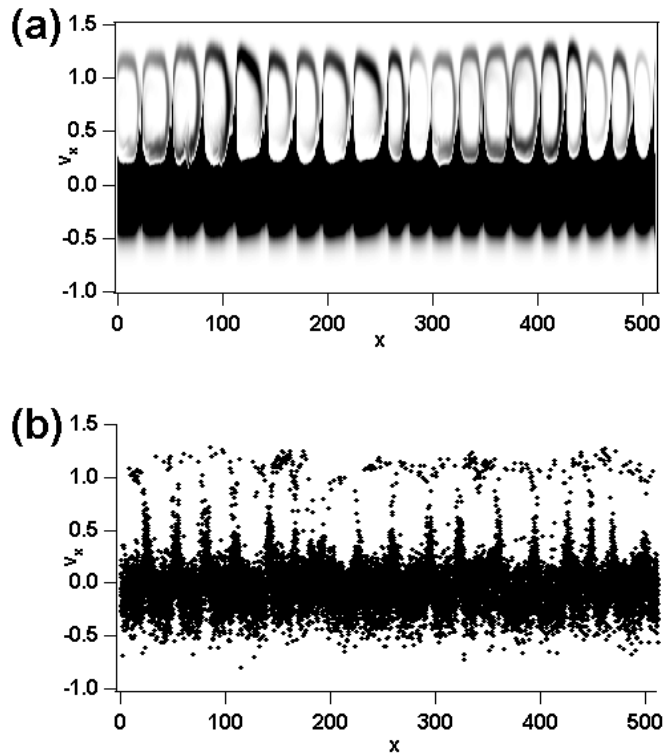}
\caption{\label{figure1} (a)The ion distribution function of the Vlasov-MHD simulation plotted in the $x$ - $v_x$ phase space at $t = 1000$ (Run 1). (b)The scatter plotts of the ion hybrid simulation plotted in the $x$ - $v_x$ phase space at $t = 620$.}
\end{figure}

We first show the simulation results (Run 1) of the parametric instability of circularly polarized Alfv\'en waves with $k_{0}=-0.4049$, $b_{0}=0.25$, and $\beta_{i}=0.07$, $\beta_{e}=0.5$, which parameters are same as those in Araneda \etal (2008). With these parameters, the modulational instabilities are dominant (Araneda \etal, 2008). Figure~\ref{figure1}(a) shows the snapshot of the ion distribution at $t = 1000$, when the instability is almost saturated. Same as the ion hybrid simulation in Araneda \etal (2008), ions trapped by the nonlinear density fluctuations are observed. Figure.~\ref{figure1}(b) shows the scatter plot of the ion hybrid simulation plotted in the $x$ - $v_{x}$ phase space at $t = 620$, which corresponds to the Fig. 4(c) in Araneda \etal (2008).

\begin{figure}
\noindent\includegraphics[width=22pc]{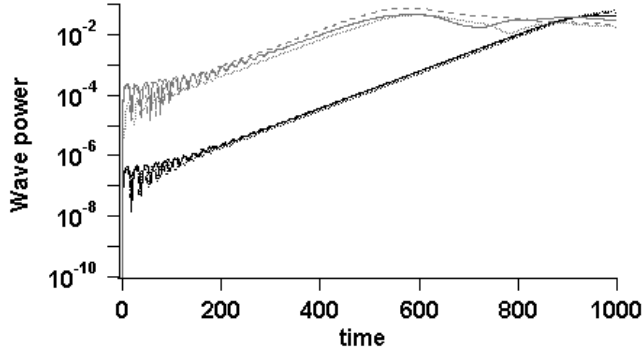}
\caption{\label{figure2} Wave power history of the wave mode with the maximum growth rates in the Vlasiv-MHD simulation (black lines) and ion hybrid simulation (gray lines). The solid, dotted and dashed lines indicate the wave power history of the wave mode with $k=-0.1593$, $k=-0.1718$, $k=-0.1847$, respectively. The growth rates (wave number) measured in the Vlasiv-MHD simulation are $0.01395$ ($k=-0.1593$), $0.01385$ ($k=-0.1718$), and $0.01388$ ($k=-0.1847$). Those measured in the ion hybrid simulation are $0.01287$ ($k=-0.1593$), $0.01352$ ($k=-0.1718$), $0.014$ ($k=-0.1847$), which are very similar to those in Araneda \etal (2008).}
\end{figure}

Figure.~\ref{figure2} shows the wave power history of the wave mode with the maximum growth rates in the Vlasiv-MHD simulation and ion hybrid simulation, respectovely. The linear growth of the Vlasiv-MHD simulation agree well with the those in the linear analysis, which maxmum growth rate is $0.0141$. Furthermore, since the initial parent Alfv\'en wave is weakly nonlinear ($b_{0}^{2}=0.0625$), the linear growths of the ion hybrid simulation agree with the linear growth of Vlasiv-MHD simulation as expected by the results in the past studies (Bugnon \etal, 2004; Nariyuki and Hada, 2007). The difference of the wave power at early time is due to the numerical noise of the ion hybrid simulation, which uses the super-particles. 

\begin{figure}
\noindent\includegraphics[width=22pc]{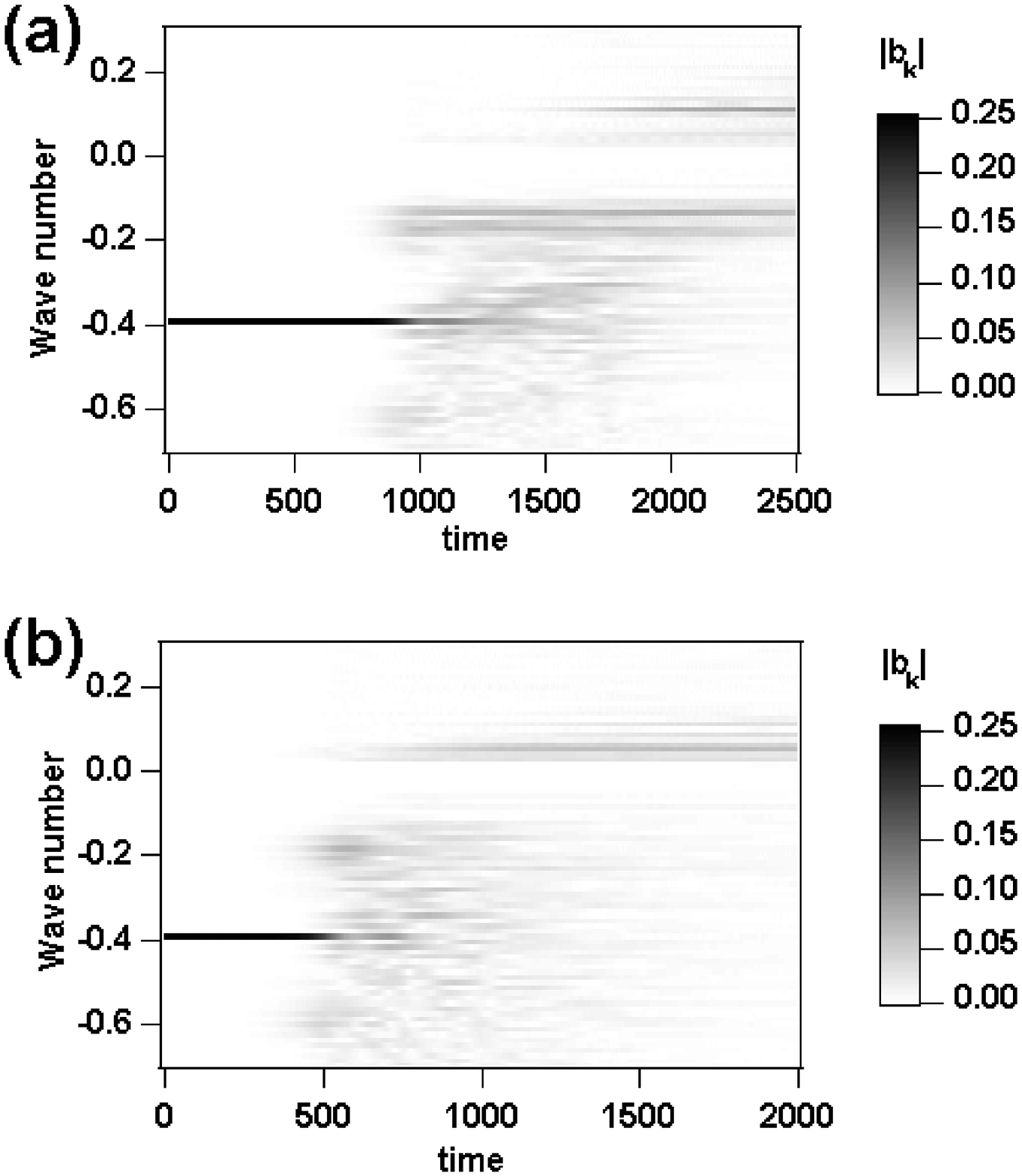}
\caption{\label{figure3} Time evolution of the magnetic field ($b_{\perp}$) power spectrum, plotted in the phase space of the wave number ($k$) and time, in (a)the Vlasov-MHD simulation and (b)the ion hybrid simulation.}
\end{figure}

Figure.~\ref{figure3} shows the time evolution of the magnetic field ($b_{\perp}$) power spectrum, plotted in the phase space of the wave number ($k$) and time, in the Vlasov-MHD simulation and the ion hybrid simulation, respectively. While the modulational instability (around $k \sim -0.2$ and $-0.6$) are very similar, the daughter waves related to the secondary decay instability (wave excitation at $k>0$) are excited at the different wave number (around $k \sim 0.1$ in the Vlasov-MHD simulation and $k \sim 0.05$ in the ion hybrid simulation). It is worth noting that while the assumptions in the Vlasov-MHD simulations are thought to be the main cause of such a difference, numerical noises in the ion hybrid simulations possibly influence the resulting plasma conditions at the nonlinear stage, since these also influence the parametric instabilities themselves. Thus, the extended Vlasov-MHD simulation and the full-Vlasov simulation are necessary to clarify the cause of such differences.

As a matter of fact, the Vlasov-MHD system corresponds exactly with the full-Vlasov system (the system of ion hybrid simulations) under the \lq\lq static approximation'' $\rho \propto |b|^{2}$ (Mj{\o}lhus, 1976; Mj{\o}lhus and Wyller, 1988), which is consistent with the condition that parallel electric field ($E_{x}$) is the potential force and the plasmas are isothermal. We here emphasize that in spite of a lot of assumptions, as far as the nonlinearlity in the system is not so large, the present Vlasov-MHD simulation model validly solves time evolution of the Alfv\'en waves both in the linear and nonlinear stage.


\begin{figure}
\noindent\includegraphics[width=22pc]{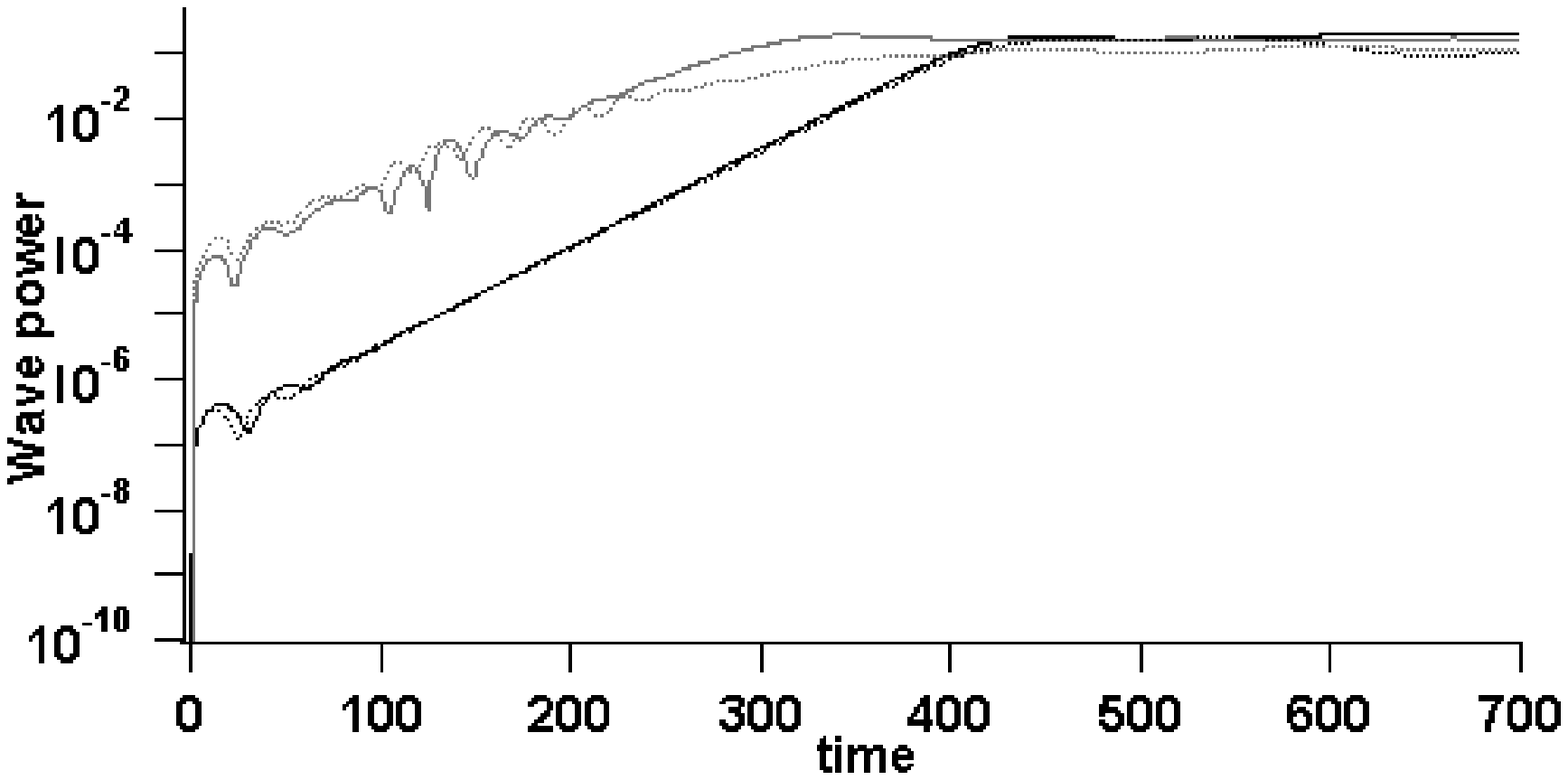}
\caption{\label{figure4} Wave power history of the wave mode with the maximum growth rates in the Vlasiv-MHD simulation (black lines) and ion hybrid simulation (gray lines). The solid and dotted lines indicate the wave power history of the wave mode with $k=-0.098172$, $k=-0.12271$, $k=-0.1847$, respectively. The growth rates (wave number) measured in the Vlasiv-MHD simulation are $0.0344$ ($k=-0.098172$) and $0.0339$ ($k=-0.12271$). Those measured in the ion hybrid simulation are $0.0262$ ($k=-0.098172$) and $\sim 0.0243$ ($k=-0.12271$).}
\end{figure}

\begin{figure}
\noindent\includegraphics[width=22pc]{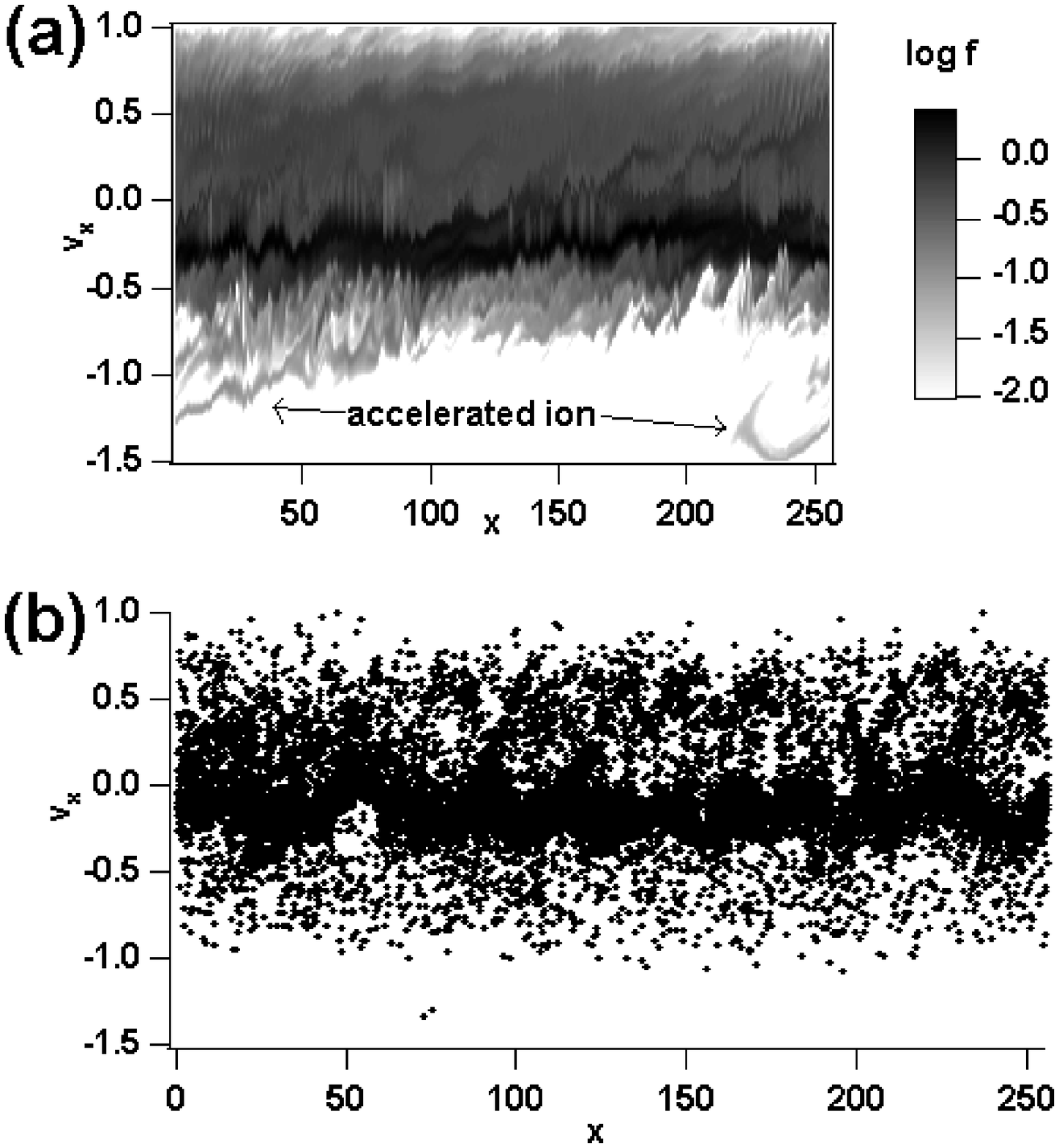}
\caption{\label{figure5} (a)The ion distribution function of the Vlasov-MHD simulation plotted in the $x$ - $v_x$ phase space at $t = 1000$ (Run 2). (b)The scatter plotts of the ion hybrid simulation plotted in the $x$ - $v_x$ phase space at $t = 800$. In the present paper, $25,000$ particles (1/200 of the total particles) are plotted.}
\end{figure}

We finally mention about the simulation results (Run 2) of the parametric instability of circularly polarized Alfv\'en waves with $k_{0}=0.196$, $b_{0}=0.4$, and $\beta_{i}=0.01$, $\beta_{e}=0.32$. The decay instability is dominant with these parameters (see Nariyuki and Hada, 2007). Figure~\ref{figure4} shows the wave power history of the wave mode with the maximum growth rates in the Vlasiv-MHD simulation (black lines) and ion hybrid simulation (gray lines). In contrast to the previous run, since the amplitude of the initial parent wave is relatively large ($b_{0}^{2}=0.16$), the linear growth rates of the Vlasov-MHD system disagree with the ion hybrid simulation (Nariyuki and Hada, 2007). At the nonlinear stage, the backward acceleration is caused by the steepen wave packets of excited daughter Alfv\'en waves in the Vlasov-MHD simulations (Fig.~\ref{figure5}(a)). We remark that such an acceleration is hardly observed in the present ion hybrid simulation (Fig.~\ref{figure5}(b)). It is because that since the number of accelerated particle is very small, the number of particles (1500 per cell in the present simulation), which is not smaller than that in the past studies, is insufficient to clarify the acceleration.

Actually, as shown in Fig.~\ref{figure4}, the results of Vlasov-MHD simulations are not valid when the amplitude of the parent Alfv\'en waves are relatively large. On the other hand, we infer that such a backward acceleration \textit{can} occur in the full Vlasov system. Of course, it does not ensure the quantitative validity of the Vlasov-MHD model. We believe that the Vlasov-MHD model can make the suggestions to the ion hybrid simulation, which usually need much more computational load than the Vlasov-MHD model due to use of the large number of particles.


\section{Conclusion and Discussion}
In the present paper, we discussed the nonlinear evolution of finite amplitude Alfv\'en waves using the Vlasov-MHD simulation code, which is a generalized Landau fluid model including the nonlinear wave-particle interactions. It was confirmed that while the Landau damping should be evaluated along perturbed field lines (Finn and Gerwin, 1996), as far as the nonlinearlity of parent Alfv\'en waves are weak, numerical results of the present Vlasov-MHD simulation agree well with those of the ion hybrid simulations both in the linear and nonlinear stage. It is worth noting that while the computational load of Vlasov-MHD code is much less than that of ion hybrid simulation. On the other hand, when the nonlinearlity of parent Alfv\'en waves are not weak, the present Vlasov-MHD simulation is not proper. Furthermore, the Vlasov-MHD model is also inadequate for the high ion beta plasmas, since the finite Larmor radius effects are neglected.

We note that in spite of restriction on nonlinearlity and ion kinetics, the present Vlasov-MHD model can be applied to several heliospheric problems. One of them is the solar coronal heating / solar wind acceleration by Alfv\'en waves propagating from the photosphere (Suzuki and Inutsuka, 2006). Since the amplitude of the magnetic fluctuations is not so large and the beta ratio is small (Suzuki and Inutsuka, 2006; Tanaka \etal, 2007), the assumption in the Vlasov-MHD system is well satisfied. 

The characteristics of the observed velocity distribution function near the sun (\eg Marsch, 2006) is one of the important constraint on the solar coronal heating and solar wind acceleration model. Namely, the generation processes of the heliospheric nonequilibrium plasmas such as ion beam components, temperature anisotropy, and the relative speed among the ion spices should be comprehensively examined in the model. We note that such a point of view is important for the collaborating works among future missions (BepiColombo, SCOPE/Cross-Scale, Solar-C, Solar Orbiter, and so on), which achievements will contribute toward the heliospheric science.


\section*{Acknowledgement}

This work was supported by Grant-in-Aid for Young Scientists (Start-up) No.20840042 (Y. N.) from JSPS and Grant-in-Aid for Young Scientists (B) No.21740352 (T. U.) from MEXT of Japan. The hybrid simulation code was kindly provided by Dr. K. Tsubouchi, and the simulation run was performed with the KDK system of the Research Institute for Sustainable Humanosphere (RISH) at Kyoto University as a collaborative research project.


\end{document}